\documentclass{aipproc}
\layoutstyle{8x11single}

\renewcommand{\d}{ \textbf{d}}

\usepackage{units}

\title{Coherent Pion Production by Neutrinos\footnote{presented by E. A. Paschos at the Sixth International Workshop on Neutrino-Nucleus Interactions in the Few-GeV Region (NuInt09), May 18-22, Sitges, Barcelona, Spain }}
\author{E. A. Paschos}{address={Department of Physics, TU Dortmund, D-44221 Dortmund, Germany}}
\author{D. Schalla}{address={Department of Physics, TU Dortmund, D-44221 Dortmund, Germany}}

\begin{document}
\begin{flushright}
DO-TH-09/16 
\end{flushright}


\begin{abstract}
In this talk I review the main features of the coherent/diffractive pion production by neutrinos on nuclei. The method is based on PCAC and relates the reaction $\textbf{boson} + \textbf{nucleus} \rightarrow \textbf{pion} + \textbf{nucleus}$ to elastic pion-nucleus scattering. Estimates for the expected rates and distributions in neutrino reactions are presented with the help of hadronic data. The absolute rates are significantly smaller than the older estimates \cite{Rein:1982pf} which brings theory in agreement with the neutral current measurements and the bounds in charged current reactions.
\end{abstract}

\maketitle

\section{Introduction}

During the past year we wrote two articles dealing with neutrino interactions \cite{Paschos:2009ag,Paschos:2008gs}. Both articles use helicity cross sections for the scattering of gauge bosons  on a target which simplifies the calculations. It became evident to me and my collaborators that earlier calculations made ad hoc assumptions which can be eliminated. Thus in two articles \cite{Paschos:2009ag,Paschos:2005km} we elaborate how PCAC can be applied effectively to coherent pion production and we defined the kinematic regions where it is valid. Then restricting the analysis and the phase space integrations to this kinematic region we obtained with the help of experimental data \cite{PhysRev.122.252,Binon:1970ye,Takahashi:1995hs} cross sections, which are much lower than previous estimates \cite{Rein:1982pf} and in good agreement with experimental values and bounds for charged and neutral current eactions. In this talk I will try to clarify the method. 

The neutral and charged current reactions are
\begin{eqnarray}
\nu_\mu (k) N (p) &\rightarrow& \nu_\mu (k') N (p') \pi^0 (p_\pi) \\
\nu_\mu (k) N (p) &\rightarrow& \mu^- (k') N (p') \pi^+ (p_\pi) 
,\end{eqnarray}
where $N$ is a nucleus with mass $M$. For the process we use variables in the rest frame of the nucleus with $q=k-k'$, $Q^2 = - q^2$, $\nu=E-E'$, $E=k_0$, $E'=k_0'$ and $t=(q-p_\pi)^2$. In the early experiments \cite{Faissner:1983ng,Marage:1984zy} the groups observed a sharp peak in the $t$-dependence of the cross section in nuclei when the nucleus remained in tact (events without stubs). These are the events that the new neutrino experiments are searching for in what it is traditionally called  "coherent pion production by neutrinos" or shortly "coherent scattering". I must mention, however, that some articles refer to the process as coherent/diffractive scattering \cite{Marage:1984zy}.

\section{Theory}

I will begin by asking under which scattering conditions the nucleus remains intact. Elastic pion-nucleus scattering provides an example with the reaction taking place at various kinematic regions. When the wavelength corresponding to the momentum transfered to the nucleus is larger than the nucleus the cross section is large. When the wavelength is smaller than the nucleus the cross section becomes very small, decreasing exponentially with the momentum transfer. Evidently the reaction takes place also for wavelengths smaller than the nucleus.

In neutrino scattering there is an interesting kinematic domain where
\begin{itemize}
 \item[(i)] $Q^2 ~$ is a few $m_\pi^2$ for PCAC to be applicable, and
 \item[(ii)] $\nu^2 >> Q^2$, where the leptonic current is dominated by the helicity-zero polarization, i.e. the leptonic current takes the divergence of the hadronic current.
\end{itemize}

Under these two conditions the contribution of the hadronic axial current is replaced by the corresponding pion-target amplitude. In other words, the symmetry of the hadronic interactions allows the replacement of the
\begin{equation}
                                    \textbf{Axial-current} + \textbf{target} \Longrightarrow X  \end{equation}
by
\begin{equation}                                               \textbf{pion} + \textbf{target} \Longrightarrow  X
\end{equation}
and guarantees smooth dependence of the amplitudes on the variables $\nu$ and $Q^2$. The replacement is guaranteed not only at a point but in a kinematic domain. 

Two more remarks are now in order:
\begin{itemize}
 \item[a)] the framework is applicable to other reactions provided conditions (i) and (ii) are satisfied,
 \item[b)] accurate calculations must include the effects from the finite mass of the muon in CC reactions.  
Muon mass corrections have been included in various approximations \cite{Adler:2005ada,Berger:2007rq} and we developed formulas where the lepton mass is included exactly.
\end{itemize}

I present next the formulas for charged and neutral current reactions
\begin{eqnarray}
\frac{\d \sigma_{CC}}{\d Q^2 \d \nu \d t} &=& \frac{G_F^2 |V_{ud}|^2}{ 2 (2\pi)^2} \frac{\nu}{E_\nu^2} \frac{f_\pi^2}{Q^2} \left\{ \tilde{L}_{00} + \tilde{L}_{ll} \left( \frac{m_\pi^2}{Q^2+m_\pi^2} \right)^2 + 2 \tilde{L}_{l0} \frac{m_\pi^2}{Q^2+m_\pi^2}  \right\} \frac{\d \sigma_\pi}{\d t}
\label{CC}  \\
\frac{\d \sigma_{NC}}{\d Q^2 \d \nu \d t} &=& \frac{G_F^2}{4 (2 \pi)^2} \frac{\nu}{E_\nu^2} \frac{f_\pi^2}{Q^2} \tilde{L}_{00} \frac{\d \sigma_\pi}{\d t}
 \label{NC}\end{eqnarray} 

with

\begin{eqnarray}
\tilde{L}_{00} &=& 2 \frac{\left[ Q^2(2E_\nu - \nu) - \nu m_\mu^2 \right]^2}{Q^2 (Q^2 + \nu^2) } - 4 (Q^2 + m_\mu^2)\label{L00}\\
\tilde{L}_{l0} &=& 2 m_\mu^2 \frac{Q^2 (2E_\nu - \nu ) - \nu m_\mu^2}{ Q^2 \sqrt{Q^2+\nu^2}} \\
\tilde{L}_{ll} &=& 2 m_\mu^2 \left( 1 + \frac{m_\mu^2}{Q^2} \right)
.\end{eqnarray}

For these formulas we assumed that an averaging over the azimuthal angle $\phi$ has taken place. The angle $\phi$ is defined between the neutrino-muon and the W-pion plane. Smaller terms from other helicities have been omitted \cite{Paschos:2005km}.

The triple differential cross sections can be compared directly with data. The recent experiments, however, sum over several variables which forces us to integrate over the variables $t$ and $\nu$. The integration over $t$ is in the interval from 
\begin{equation} | t_{min} | =  \left( \frac{Q^2+m_\pi^2}{2\nu} \right)^2 \end{equation}
up to the first diffractive minimum.

For the integration over $t$ we use the data of \cite{PhysRev.122.252,Binon:1970ye,Takahashi:1995hs} in order to obtain the function 
\begin{equation} \sigma_\pi (Q^2,\nu) = \int_{|t_{min}|}^\infty \frac{\d \sigma_\pi}{\d t} \d t.\end{equation}

\section{results}

The result of the numerical integration is shown in figure 1. This approach of using experimental data was introduced in \cite{Paschos:2005km} and has now been also adopted by Berger and Sehgal \cite{Berger:2008xs} who use phase shift results from other experiments. It is important that $\sigma (Q^2, \nu)$ decreases very fast with $Q^2$ , so that the main contribution is limited to values of $Q^2$ smaller than $0.15$ or $0.20\unit{GeV^2}$, which is the region where PCAC is applicable.

\begin{figure}
 \includegraphics{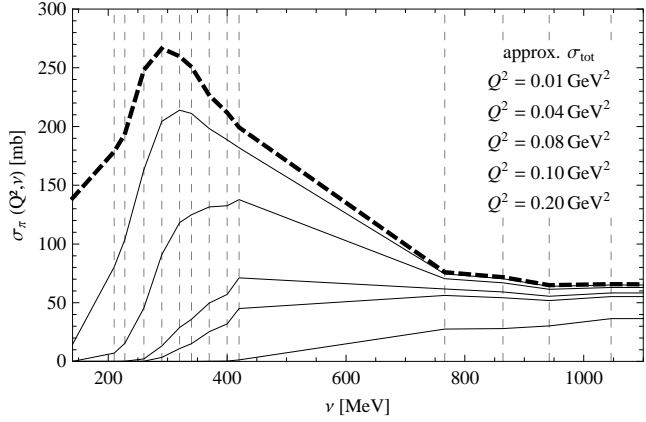}
\caption{Integrated elastic pion nucleus cross section for some momentum transfers}
\end{figure}

The integration over $\nu$ must respect the condition $\nu >> \sqrt{Q^2}$. To satisfy this condition we introduced the variable $\xi$ in the equation 
\begin{equation}                         \nu= \xi \sqrt{Q^2} \end{equation}
and selected for the $\nu$ integration the range 
\begin{equation} \max \left( \xi \sqrt{Q^2} , \nu_{min} \right) < \nu < \nu_{max} .\end{equation}

The variable $\xi$ describes a kinematic cut to be introduced in the data. We show in figures 2 the double differential cross section for $E_\nu= 1.0$ and 10.0$\unit{GeV}$ respectively. These figures show where the large values of the cross section are located. It is evident that the cross section is large for $Q^2 < 0.15 \unit{GeV^2}$ and for $1.0 < \xi < 3.0$ or $4.0$. 

\begin{figure}
 \includegraphics[width=.5\textwidth]{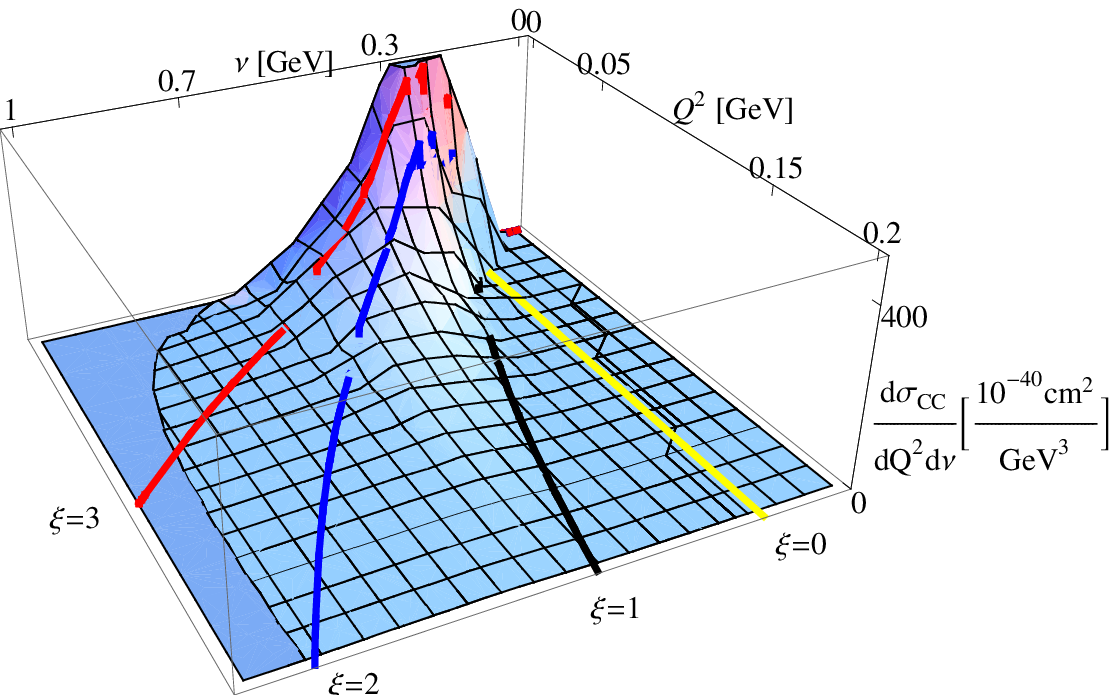} \includegraphics[width=.5\textwidth]{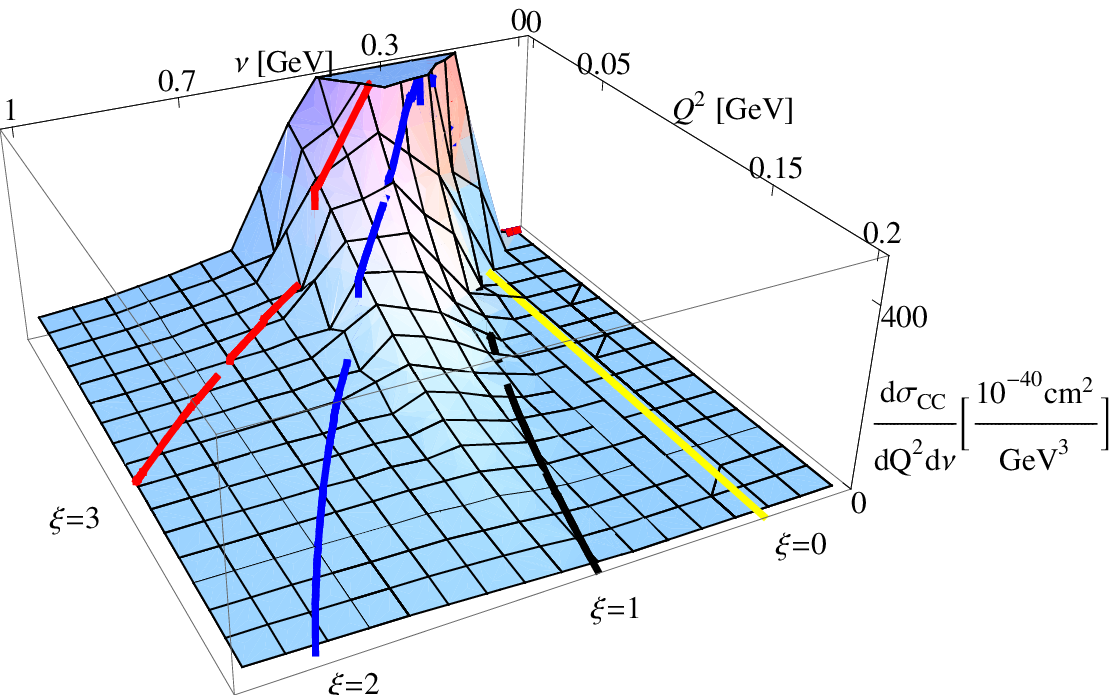}
\caption{Double differential charged current cross section at $E_\nu=1\unit{GeV}$ (left) and $E_\nu=10\unit{GeV}$ (right). The lines represent different integration limits (see text).}
\end{figure}

One may argue that the good features of the results are limited to the contribution from PCAC and other contributions may also be present. This is a valid remark which motivated us to search and estimate other contributions but found them to be very small. For instance, for neutral currents the density matrix elements $\tilde{L}_{00}$, $\tilde{L}_{RR}$, $\tilde{L}_{LL}$ are propotional to $Q^2$ and as $Q^2 \rightarrow 0$ the only helicity cross section that survives is $\sigma_s$ which has a $1 / Q^2$ dependence. For charged current reactions there is a similar argument. In this case as $Q^2 \rightarrow m_\mu^2$ only $\sigma_s \approx 1 / m_\mu^2$ survives. Contributions from other reactions have been estimated \cite{Paschos:2005km} and found to be much smaller (see the estimates between Eqs (18) and (24) of \cite{Paschos:2005km}).

The calculated differential cross sections $\d \sigma / \d Q^2$ for $E_\nu = 1.0 \unit{GeV}$  are shown in figure 3. The four curves correspond to the minimum value for the energy $\nu$ with $\xi=3$ for the lower curve, $\xi=2$ for the middle curve and $\xi= 1$ or zero for the upper curve. We note that the cross section for the two cases $\xi= 1$ or zero are almost identical, reflecting the structure observed in figure 3. For higher neutrino energies the cross section extends to larger values of $Q^2$.

\begin{figure}
 \includegraphics[width=.5\textwidth]{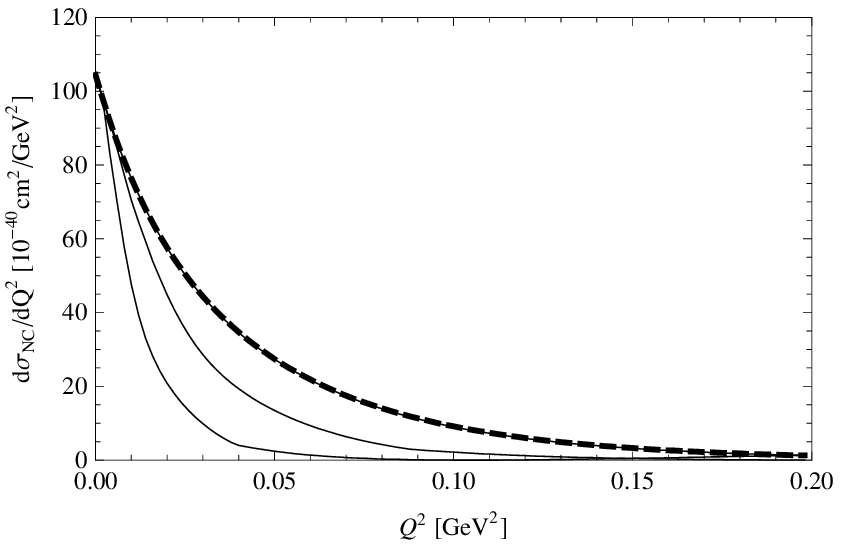}
 \includegraphics[width=.5\textwidth]{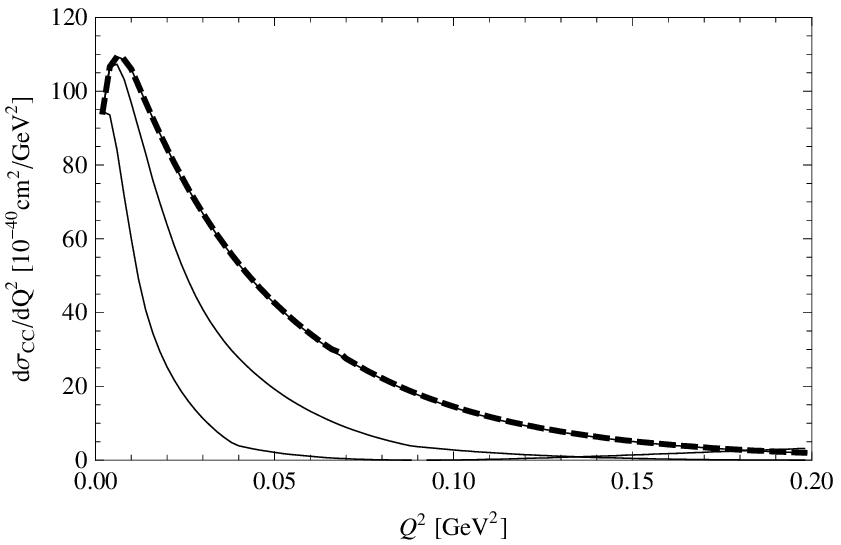}
\caption{Neutral (left) and charged (right) current differential cross sections at $E_\nu = 1.0 \unit{GeV}$.}
\end{figure}

Comparisons of the predictions in figures 2 - 4 with experimental data will verify or falsify through equations (5) and (6) the direct relation between neutrino and pion-carbon scatering. A typical signature is the concentration of events at small $Q^2$ and $1< \xi < 3$. The early experiments observed a sharp peak in the $t$-dependence \cite{Faissner:1983ng,Marage:1984zy}. The recent experiments do not have this capability and integrate over $t$ and other variables. In fact they estimate every known reaction and look for an excess in kinematic regions where coherent scattering is large. The situation is even more complicated because their theoretical estimate for "`coherent pion production"' uses the old estimate \cite{Rein:1982pf} which is $\sim 2$ times larger than the recent values \cite{Paschos:2009ag,Paschos:2005km}. For neutral current reactions there is evidence for coherent scattering \cite{Faissner:1983ng,Raaf:2005up,Isiksal:1984vh,Grabosch:1985mt}. For the charged current reactions the situation is more complicated because the background for neutrino reactions is even larger, i.e. the production of $\pi^+:\pi^0:\pi^-$ in the delta region is 9:2:1. Under these difficulties two experimental groups report upper bounds for  charged current reactions \cite{Hasegawa:2005td,Hiraide:2008eu}. For antineutrino charged current reactions the background for the production of $\pi^-$ through the delta resonance is smaller. Dr. Tanaka \cite{Tanaka} at this conference reported preliminary evidence for coherent pion production by antineutrinos (charged currents).

This situation forces us to integrate over the variable $Q^2$ as well. The theoretical framework that I described is valid in the small $Q^2$ region. At higher energies $\d \sigma / \d Q^2$ has a tail extending beyond $Q^2= 0.20 \unit{GeV^2}$ where the application of the PCAC relation is questionable. For this reason we introduced a cut in $Q^2$ and present the results in figures 4. The dependence of the integrated cross sections on the cut-off for $E_\nu<4\unit{GeV}$ is small. This follows again from the fact that the cross section is  large at small values of $Q^2$ (see figure 2). In the figures we included experimental results from several groups \cite{Faissner:1983ng,Raaf:2005up,Isiksal:1984vh,Grabosch:1985mt,Hasegawa:2005td,Hiraide:2008eu}.

\begin{figure}
 \includegraphics[width=.5\textwidth]{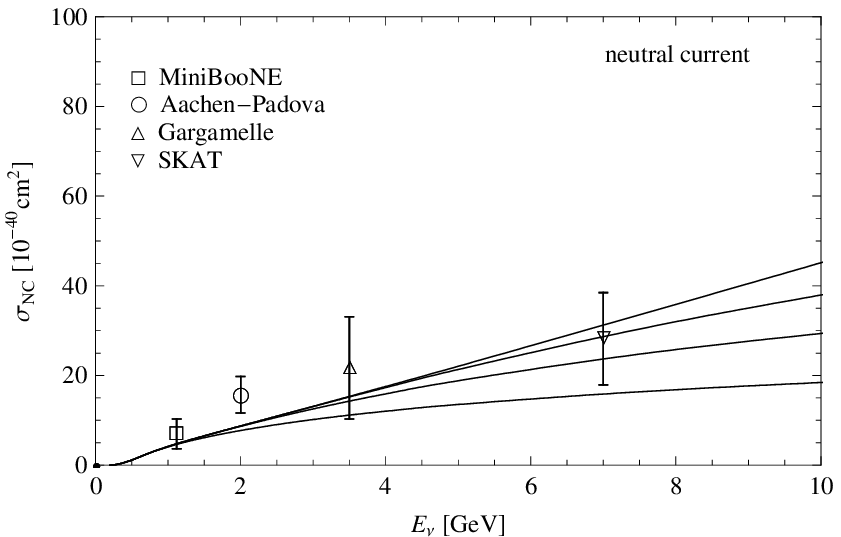}
 \includegraphics[width=.5\textwidth]{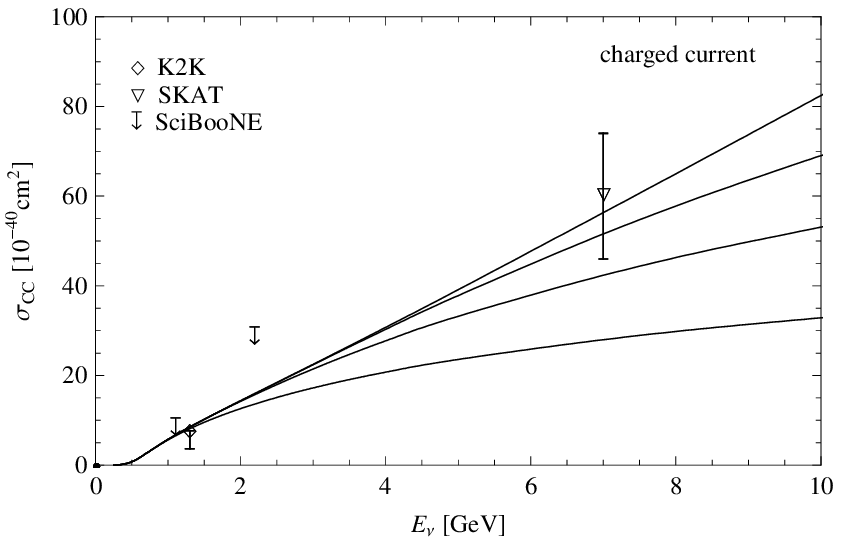}
\caption{Neutral (left) and charged (right) current integrated cross sections.}
\end{figure}

\section{Summary}
\label{sec:Summary}
The coherent/diffractive pion production by neutrinos is determined by PCAC in specific kinematic regions described at the beginning of this article. The cross sections are predicted as functions of $Q^2$, $\nu$ and $t$. When the experimental colleagues introduce the appropriate kinematic cuts, it will be possible to check the cross sections in equations (5) and (6).

Discrepancies in earlier comparisons of theory with experiment was due to the inappropriate modeling of pion-nucleus scattering \cite{Rein:1982pf}. This has been now rectified \cite{Paschos:2009ag,Paschos:2005km,Berger:2008xs} and the agreement with the experimental points (neutral current) and bounds in charged currents is good (see figure 4). Furthermore, the ratio of charged to neutral current reactions is a consequence of isospin symmetry and standard kinematics. Thus the observation of the process in the neutral current reaction determines the corresponding reaction in the charged currents.

It will be helpful to analyse data using the new formulas in order to see if an excess over and above well known reactions is present. Such an analysis has been presented by Tanaka \cite{Tanaka} for antineutrino reactions where the background from $\pi^-$ is relatively small.

\end{document}